\documentclass[%
 reprint,
superscriptaddress,
 amsmath,amssymb,
 aps,
]{revtex4-2}

\usepackage{graphicx}
\usepackage{dcolumn}
\usepackage{bm}

\begin{document}

\preprint{APS/123-QED}

\title{Nanometric Turing Patterns: Morphogenesis of a Bismuth Monolayer}
\author{Yuki Fuseya$^*$}
\affiliation{Department of Engineering Science, University of Electro-Communications, Chofu, Tokyo 182-8585, Japan}

\author{Hiroyasu Katsuno}
\affiliation{Institute of Low Temperature Science, Hokkaido University, Kita-19, Nishi-8, Kita-ku, Sapporo, Japan}

\author{Kamran Behnia}
\affiliation{Laboratoire Physique et Etude de Mat\'eriaux (CNRS-Sorbonne Universit\'e), ESPCI Paris, PSL Research University, 75005 Paris, France}

\author{Aharon Kapitulnik$^*$}
\affiliation{Departments of Physics and Applied Physics, Stanford University, Stanford, CA 94305,
United States of America}

\date{\today}

\begin{abstract}
Turing's reaction-diffusion theory of morphogenesis has been very successful in understanding macroscopic patterns within complex objects ranging from biological systems to sand dunes. However, this mechanism was never tested against patterns that emerge at the atomic scale, where the basic ingredients are subject to constraints imposed by quantum mechanics. Here we report evidence of a Turing pattern that appears in a strained atomic bismuth monolayer assembling on the surface of NbSe$_2$ subject to interatomic interactions and respective kinetics. The narrow range of microscopic parameters reflected in numerical analysis that observe stripe patterns and domain walls with Y-shaped junctions is a direct consequence of the quantum-mechanically allowed bond-lengths and bond-angles. This is therefore the first demonstration of a dynamically formed Turing pattern at the atomic scale.
\end{abstract}

\maketitle


 \begin{figure*}[tb]
 \begin{center}
 \includegraphics[width=16cm]{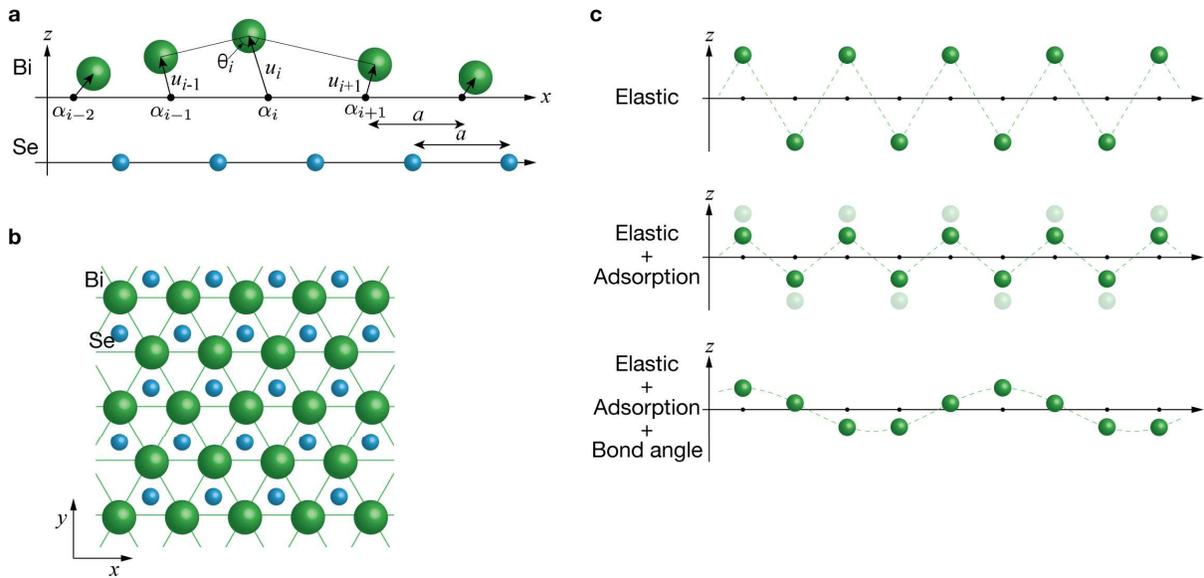}%
 \caption{\label{fig1} Crystal structure of monolayer Bi on the top Se layer. {\bf (a)} Side view and {\bf (b)} Top view. $\alpha_i$ and $\bm{u}_i$ are the positions corresponding to the adsorption potential minimum and the displacement of the Bi atom at site $i$, respectively. 
 {\bf (c)} Schematic side view of the equilibrium state. When only the elastic potential $V_e$ is considered, the Bi atoms form staggered patterns with the period $2a$. The adsorption potential $V_a$ reduces the amplitude of the staggered patterns. When the bond angle potential $V_b$ is included, the Bi atoms form patterns with wavelengths longer than $2a$.}
 \end{center}
 \end{figure*}

Ordered patterns can arise out of randomness during morphogenesis.  An explanation for this puzzle was put forward by Turing, who explained the emergence of stationary patterns by invoking interplay between an activator and an inhibitor with different diffusion rates \cite{Turing1952}. Turing's theory of diffusion-reaction has proven extremely influential across many disciplines \cite{Meinhardt_book,Koch1994,Murray_book}. For example, the pigment patterns on sea shells \cite{Meinhardt_book2}, the stripes on tropical fish \cite{Kondo1995,Kondo2010}, and the purely chemical system of chlorite-iodide-malonic acid \cite{Castets1990,Ouyang1991,Horvath2009} have been studied as Turing patterns. The typical length scale for biological pigment-based patterns ranges from millimeters to centimeters, and that for purely chemical systems is of the order of sub-millimeters. However, it has never been employed to describe atomic scale patterns. Because quantum mechanical considerations are expected to play a key role at the scale of interatomic distances, the relevance of Turing's theory in such a context is yet to be demonstrated. Here, we show that atomic monolayer bismuth grown by molecular beam epitaxy (MBE) provides such an opportunity. 

Strained MBE has become an important technique for its direct impact on various functional properties for potential applications in electronic devices, quantum sensors, or spintronics applications to name a few. Traditionally, epitaxial growth of two-dimensional films has been described within the framework of first-order phase transition, yielding three modes of epitaxial growth \cite{Bauer1958}: island growth \cite{Volmer1926}, layer growth \cite{Frank1949} and their combination, namely islands growing on one or two already-completed monolayers \cite{Stranski1937}. When equilibrium conditions are met, the morphology of the newly grown layer is obtained by balancing the elastic energy against the surface energy. However, under far-from-equilibrium conditions, kinetic processes must dominate, which arise from the relevant thermodynamic driving forces. This is the regime where predictability is poor, since small changes in initial and/or boundary conditions may have dramatic effects on the resulting morphology of the film. In particular, where the growth of metallic (or semi-metallic) films also involve quantum mechanics considerations associated with the itinerant electronic states, constraints on allowed states of bonding, imposing specific bond length or bond angle have to be met.

Our study was motivated by the recent work of Fang {\it et al.}, who first found a mysterious pattern of atomic-monolayer grown by MBE \cite{Fang2018}. The Bi monolayer on NbSe$_2$ was originally fabricated in the interests of research in two-dimensional topological physics, to which intensive efforts have been devoted in recent years \cite{Hasan2010,Qi2011,Reis2017}. The patterns in Bi monolayer exhibit stripes with a period of five atoms ($=1.7$ nm) and domain walls with Y-shaped junctions (cf. Fig. 2b), which break the symmetry of the underlying lattice \cite{Fang2018}. This mysterious pattern strikingly resembles the stripes with a Y-shaped pattern in angelfish {\it Pomacanthus}, where Turing's mechanism has been studied \cite{Kondo1995,Kondo2010}, although their length scales differ by more than six orders of magnitude.

The necessary conditions for the formation of Turing patterns are the following: (i) auto-catalysis and cross-catalysis must exist between an activator ($u_a$) and an inhibitor ($u_h$), and (ii) the diffusion of the inhibitor ($D_h$) must be much faster than that of the activator ($D_a$), i.e., $D_a \ll D_h$ \cite{Turing1952,Meinhardt_book,Koch1994,Murray_book,Kondo2010}. Turing expressed this mechanism of pattern formation in terms of a simple simultaneous differential equation of the form \cite{Turing1952}
\begin{align}
	\frac{\partial u_{a, h}}{\partial t}=D_{a, h} \nabla^2 u_{a, h} + f_{a, h}(u_a, u_h),
	\label{RD}
\end{align}
which is called the reaction-diffusion equation. ($f_{a, h}(u_a, u_h)$ are reaction functions.)
In principle, no upper or lower limit of the length scale is imposed by the reaction-diffusion equation. In other words, Turing patterns possess an intrinsic wavelength, which depends only on the ratio of the parameters in the equation. This scaleless property of Turing patterns contrasts with the other well-known non-equilibrium  structures, such as the convective B\'enard cells and the Taylor vortices in Couette flows, the length scales of which are restricted by the geometric scales of the system \cite{Chandrasekhar_book,Castets1990}. Therefore, one naively expects that Turing patterns can be realized at any arbitrary length scale. However, the length scale of patterns hitherto documented in biology and chemistry do not fall below a fraction of a millimeter \cite{Koch1994,Kondo2010,Castets1990,Ouyang1991,Horvath2009}. It is of fundamental interest to determine the validity of Turing theory at the interatomic scale, where the solid state remains stable thanks to quantum mechanics \cite{Dyson1967}. If Turing patterns can be formed even at atomic scales, what are the activator and the inhibitor? Why are their diffusion rates so different? What determines the wavelength? We provide answers to each of these questions in the following discussion.

The present work aims to elucidate the origin and underlying mechanism of the mysterious pattern exhibited by Bi monolayers on NbSe$_2$. The lattice constant of the Bi monolayer on NbSe$_2$ matches that of the hexagonal lattice in the NbSe$_2$ ($a = 3.45$ \AA) \cite{Fang2018}. This is substantially shorter than the lattice constant of bulk Bi, $b = 4.53 $ \AA\,, or of the Bi layer in Bi$_2$Se$_3$, $b = 4.13$ \AA. Therefore, it can be expected that the NbSe$_2$ substrate generates substantial lateral strain in the Bi monolayer, inducing ripples to relieve the strain. However, the ripples should maintain the rotational symmetry of the hexagonal lattice because the direction of strain is three-fold. The expected ripple pattern due to strain would be completely different from the mysterious pattern observed in the Bi monolayer.

The crystal structure of Bi monolayer on NbSe$_2$ is depicted in Fig. 1. Bi atoms are deposited on the top Se layer of the substrate, and they form a hexagonal lattice with a lattice constant equal to that of the Se layer. Let us consider empirical two-body interatomic potentials between Bi--Bi atoms (elastic potential $V_e$) and Bi--Se atoms (adsorption potential $V_a$). Although the potentials are generally given by the Lennard-Jones or Morse potentials, it is sufficient to consider the potential around its minimum in the present case, namely, these are well approximated by the harmonic oscillator. The elastic potential is given by $V_e =(K_e/2)\sum_{ij}(b -\ell_{ij})^2$, where $K_e$ is the elastic constant, and $b$ ($>a$) is the natural lattice constant of Bi. $\ell_{ij}=|\bm{u}_i-\bm{u}_j + \bm{\alpha}_{ij}|$ is the distance between the $i$-th and $j$-th Bi atoms, $\bm{\alpha}_{ij}$ the vector between the $i$-th and $j$-th sites, and $\bm{u}_i$ is the displacement of Bi atoms at the $i$-th site. The Se atoms are tightly bound to the rest of the NbSe$_2$ substrate; consequently, the lattice distortion of Se is negligibly small. Therefore, the adsorption potential can be expressed as a one-body potential of the form $V_a = (K_a/2)\sum_i |\bm{u}_i|^2$.

If Bi atoms interact only through these two-body potentials, they would simply form a staggered pattern to release the elastic energy, as shown in Fig. 1c. This clearly fails to explain the observed patterns. This failure is due to the lack of consideration of covalency. In particular, the covalency is known to be crucial for understanding the crystal structure of pnictogens (Group 15), even if they are semimetals \cite{Littlewood1980,Behnia2016}. To consider the covalency of the crystal, it is necessary to include the three-body potential \cite{Tersoff1988}, which can be expressed in terms of the bond angle in Bi--Bi--Bi. (The bond angle term corresponding to Se--Bi--Se is renormalized to the one-body term $V_a$ because the position of the Se atoms is fixed.) The bond angle potential is given by $V_b=K_b a^2 \sum_i (1+\cos \theta_i)$, where $K_b$ is the coefficient that has the same dimension as $K_{e, a}$. Further, $\theta_i$ is the bond angle corresponding to Bi--Bi--Bi, and $i$ denotes the center Bi atom (Fig. 1a). This bond angle potential is crucial in obtaining the Turing patterns. 

 \begin{figure}[tb]
 \begin{center}
 \includegraphics[width=8cm]{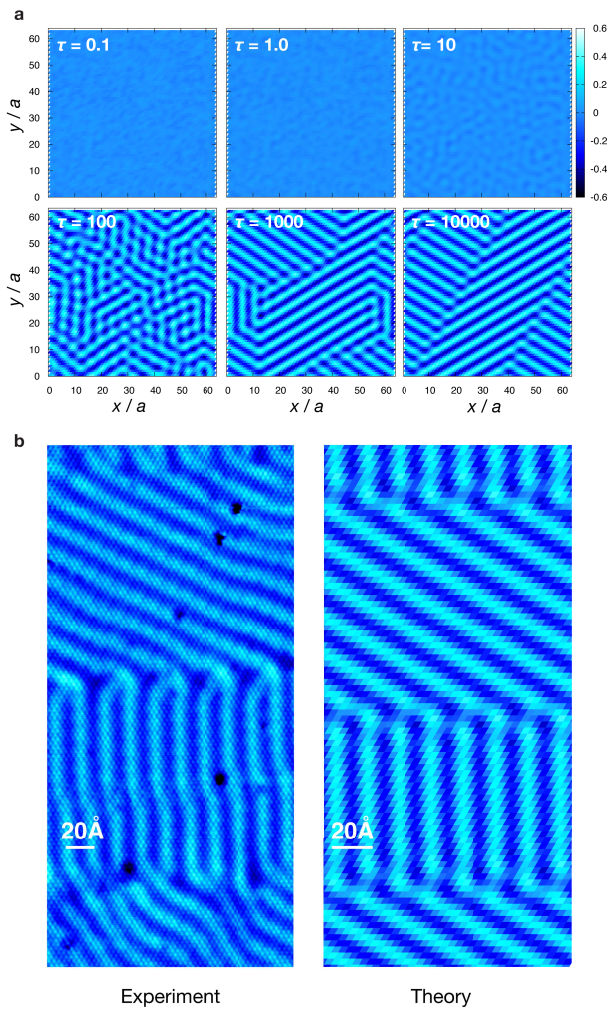}%
 \caption{\label{fig2} {\bf (a)} Time evolution of the vertical displacement $u_z$ in the $64 \times 64$ hexagonal lattice with a periodic boundary condition for $b/a=1.3$, $K_a/K_e=0.3$, and $K_b/K_e=0.12$. The unit of time is given by $K_e^{-1}$.
 {\bf (b)} Comparison of the numerical simulation at $\tau=10000$ and the experiment by Fang {\it et al.} \cite{Fang2018}. The wavelength obtained via the simulation is $5a$ ($=1.7$ nm), which is in excellent agreement with the experiment.}
 \end{center}
 \end{figure}

We investigated the dynamics of this model. Generally, a non-conserved quantity will evolve such that the total energy decreases. In particular, we aim to analyze the dynamics of the displacement $\bm{u}_i$ of Bi atoms. The time evolution equation is given as $\partial \bm{u}_i /\partial t = -\eta \partial V_{\rm tot}/\partial \bm{u}_i$, where $V_{\rm tot}=V_a + V_e + V_b$ and $\eta >0$ is a prefactor. (Hereinafter, we use the renormalized time $\tau =\eta t$, whose unit is $K_e^{-1}$.) This is a simultaneous differential equation with respect to $u_i^{x, y, z}$. The explicit form of the time evolution equation is provided in the Supplementary Information \cite{SM}.
Figure 2a shows the time evolution of the vertical displacement $u_i^z$. The initial condition is given as a random distribution of $\bm{u}_i$ with a small amplitude. Notably, the system spontaneously breaks the symmetry and a pattern with a unique wavelength is formed with time. We observed that stable stripe patterns with domains are formed, with parameters $K_a/K_e=1.2$, $K_b/K_e=0.3$, and $b/a=1.3$. The period (wavelength) of the stripes is $5a$, which is in exact agreement with the experiment. The domains are oriented such that they form an angle of 120$^\circ$ with each other, and the aligned Y-junctions clearly appear at the domain boundaries. All of these theoretical results are in excellent agreement with the experimental results previously obtained with monolayer Bi on NbSe$_2$ (Fig. 2b). The maximum vertical displacement is $\sim 0.5a$, which is larger than that obtained in the experiment to some extent. If different initial conditions are employed, stripe patterns with a different direction and domain structure but the same wavelength are obtained \cite{SM}.

In the following, we will demonstrate that our time evolution equation is equivalent to Turing's reaction-diffusion equation, although they may initially seem entirely different. (See \cite{SM} for a detailed derivation.)
The vertical displacement $u_i^z$ corresponds to the activator, and the horizontal displacement $u_i^{x, y}$ corresponds to the inhibitor. The bond angle potential can be simplified as $- K_b \nabla^4 u_i$, which entails dominant (long-range) diffusion with regard to all $u_i^{x, y, z}$ \cite{Murray_book}. 
By contrast, the elastic potential gives $-K_e b/a \nabla^2 u_i^z$, which plays a role of auto-catalysis (a positive feedback) and reduces the diffusion only in the $z$-direction, resulting in $D_a \ll D_h$. Further, the elastic potential contributing to the reaction terms $f_{\mu}$ consists of $\mathcal{O}(u_i^2)$, which generates the cross-catalysis. The adsorption term reduces the wavelength of the patterns, which is crucial for achieving quantitative agreements with the experiments. 
Finally, our time evolution equation can be simplified as
\begin{align}
	\frac{\partial u_i^\mu}{\partial \tau}&= D_\mu \widetilde{\nabla}^2 u_i^\mu - K_a u_i^\mu +f_\mu (u_i^z, u_i^x),
	\label{eq1}
\end{align}
where $\widetilde{\nabla}=a\nabla$, $\mu=z, x$, $D_z = D_x-K_e b/a$, and $D_x=-K_b \widetilde{\nabla}^2 +K_e -2K_b$. (Here, we omitted the $y$-components to make the argument as clear as possible. However, the essence of our model is fully accounted for this one-dimensional array.) 
This is essentially equivalent to the reaction-diffusion equation \eqref{RD}. 
Therefore, the excellent agreement between the theory and the experiment strongly suggests that the unique patterns appearing in monolayer Bi are realized by the Turing mechanism.
To the best of our knowledge, a wavelength of 2 nm is the shortest length scale reported for Turing patterns thus far. At the same time, this agreement is strong evidence that Turing's theory is valid from centimeters to nanometers, a surprisingly wide extent.

 \begin{figure}[tb]
 \begin{center}
 \includegraphics[width=7cm]{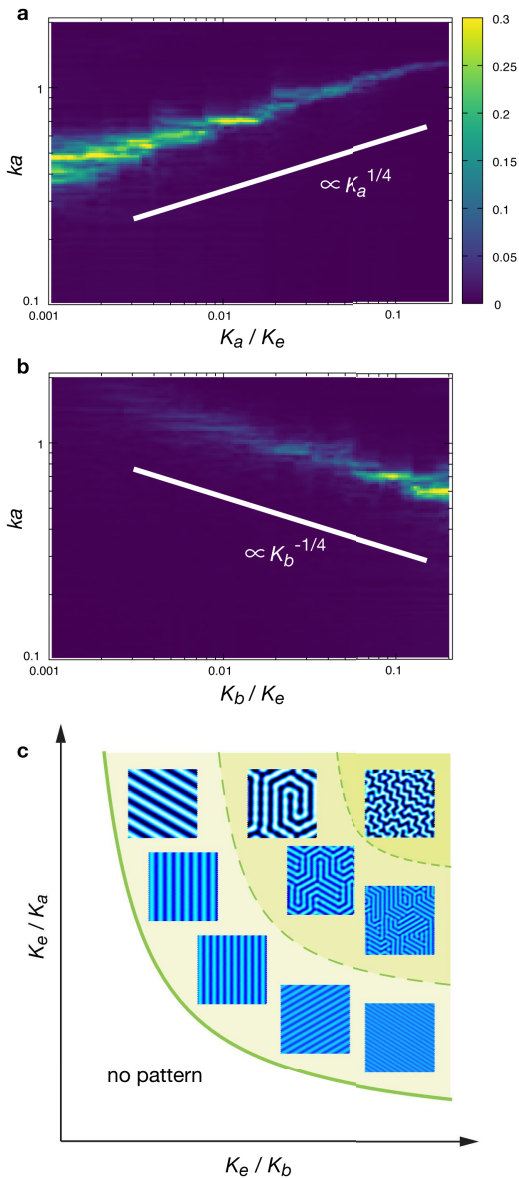}%
 \caption{\label{fig3} {\bf (a) \& (b)} Fourier component $c_k$ of the numerically obtained patterns in the model of equation \eqref{eq1} for 1024 sites and $b/a=1.1$ with (a) $K_b/K_e=0.1$ and (b) $K_a/K_e=0.01$. The brightest region corresponds to the most dominant $c_k$ of the patterns, which clearly exhibit $(K_a/K_b)^{1/4}$ dependencies.
  {\bf (c)} Pattern diagram as a function of $1/K_a$ and $1/K_b$. The Turing pattern appears in the region corresponding to $K_a K_b < \beta K_e^2$.}
 \end{center}
 \end{figure}

The most unstable wavenumber of equation \eqref{eq1} is given by
\begin{align}
	k_m = a^{-1}\sqrt{k_+ k_-} =a^{-1}\left( K_a/K_b \right)^{1/4},
	\label{km}
\end{align}
according to linear stability analysis \cite{Murray_book,SM}. This form clearly indicates that the wavelengths are determined only by intrinsic parameters.
The bond angle potential enhances the pattern wavelength, whereas the adsorption potential reduces it.
Figure 3a and b shows plots of the Fourier component $c_k$ of the patterns obtained by our numerical simulation with the non-linear terms $f_\mu (u_i^z, u_i^x)$, equation \eqref{eq1}, as a function of $K_a$ and $K_b$. The most dominant $c_k$ (the brightest region) exhibits a clear $(K_a/K_b)^{1/4}$ dependence, which agrees well with the analytic result of equation \eqref{km}. The $(K_a/K_b)^{1/4}$ dependence never changes even for the different system sizes \cite{SM}, indicating the intrinsic nature of the wavelength that is inherent in Turing patterns.

In addition to the wavelength, the pattern can be modified by tuning the parameters. The changes in the pattern are schematically depicted in Fig. 3c. (The precise pattern diagram is given in \cite{SM}.) The patterns are formed only for $K_a K_b < \beta K_e^2$, where the prefactor $\beta$ depends on $b/a$. 
Otherwise, Bi atoms form a perfectly flat surface with a hexagonal lattice even if the initial film is bumpy. The stripe patterns appear in the narrow region close to the boundary.
From these pattern diagrams, we infer the reason for the formation of the Turing patterns in Bi monolayers as follows: First, it is well known that  bulk Bi exhibits substantial structural instability owing to its specific electronic states \cite{Peierls_book2,Hoffmann_book,Fuseya2015}. In fact, Bi under pressure has 10 different solid phases \cite{Young1975}. (Only sulfur has more phases among the elemental solids \cite{Young1975}.) This structural instability is further enhanced in ultrathin Bi, resulting in a rich variety in the structure \cite{Nagao2004,Bollmann2011,Reis2017,Fang2018}. The structure instability reduces the bond angle potential, i.e., reduces $K_b$. In typical materials, $K_b$ is substantially large, and the system corresponds to the ``no pattern" region. By contrast, in the case of Bi, $K_b$ is relatively low; consequently, the system crosses the boundary and enters the pattern formation region. 

 \begin{figure}[tb]
 \begin{center}
 \includegraphics[width=8cm]{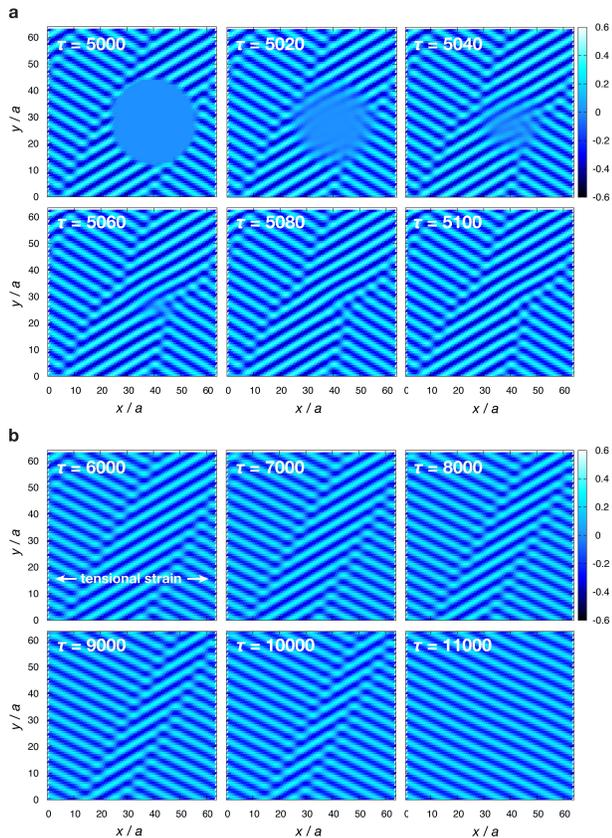}%
 \caption{\label{fig4} {\bf (a)} Simulation of wound healing. The patterns are wounded, $\bm{u}_i=0$, at $\tau=5000$.
{\bf (b)} Simulation of uniaxial stress. The lattice constant along the $x$-direction is extended by 3\% at $\tau=5000$ to simulate the application of uniaxial tensional strain. In both simulations, the same parameters as in Fig. 2 are used.}
 \end{center}
 \end{figure}

Turing patterns are generated by interference between nonlinear waves. They appear static; however, they are in dynamic equilibrium. One of the significant consequences of the dynamic property of Turing patterns is wound healing \cite{Kondo1995,Kondo2010,Murray_book}. Particularly, wound healing is crucial for living creatures to support their lives, and it can be explained using reaction-diffusion equations \cite{Murray_book}. Herein, we show that inorganic solids also possess the ability of wound healing. Figure 4a shows the wound healing property of our model. We inflicted a ``wound" by imposing $\bm{u}_i =0$ in a circular area at $\tau = 5000$. The wound rapidly healed before $\tau=5100$. Surprisingly, the domain structures and Y-junctions are regenerated exactly as before at approximately $\tau=6000$ (shown in \cite{SM}). This clearly indicates the dynamic nature of the Turing patterns generated by our model. 

We show further dynamic properties of the patterns by applying uniaxial strain. We simulated uniaxial strain by changing the lattice constant along one particular direction. We applied tensional strain (increased the lattice constant by 3\%) along the $x$-direction after $\tau = 5000$. Notably, as shown in Fig. 4b, the two separated domains combined even though the tensional strain was applied in the direction that would lead to their further separation. Consequently, the domain structure of the patterns can be controlled by applying strain.

Unnoticed Turing patterns can be ubiquitously detected in solid-state physics publications. For example, magnetic patterns in ferrimagnetic garnet films (BiGdY)$_3$(FeGa)$_5$O$_{12}$ \cite{Kudo2007} may be reinterpreted as Turing patterns. The patterns exhibited by type-I superconductors in the intermediate state \cite{Jeudy2006,Prozorov2007} or those exhibited by ferromagnetic superconductors \cite{Paulsen2012} may have a close relationship to Turing patterns because the normal or ferromagnetic states inhibit the superconducting states of the activator. The common concept of spontaneous symmetry-breaking due to diffusion-driven instability hidden in these patterns will bring about a new point of view in solid-state physics. 

Our reaction-diffusion equation implies more than the fact that it can explain the mysterious pattern of Bi on NbSe$_2$. It proposes that the patterns can be controlled by changing parameters, i.e., changing substate, overlayer, or growth conditions. It can even remove undesirable patterns and make perfectly flat thin films. Obtained different patterns can be building blocks to new devices or new physics that was prior unexpected. It enables us to design structures with long-periods, which create new band structures like Moir\'e bands in twisted bilayer graphene \cite{Bistritzer2011,Cao2018}.
We believe that this approach, together with recent advances in machine learning protocols, may pave the way for the design of new materials that may otherwise not be predicted by equilibrium thermodynamics.

\begin{acknowledgments}
We thank A. Fang for providing the raw STM data discussed in this manuscript, and N. Sasaki and K. Izawa for fruitful discussions. 
Funding: This work was initiated through a “QuantEmX” Exchange Awards (AK) and the Promotion of Joint International Research (YF) at ESPCI. Work at UEC (Japan) was supported by the JSPS Grant No. 15KK0155 and 19K21844. Work at Stanford University was supported by the U. S. Department of Energy (DOE) Office of Basic Energy Science, Division of Materials Science and Engineering at Stanford under contract No. DE-AC02- 76SF00515.
\end{acknowledgments}

\section*{Author contributions}
Y.F., K.B., and A.K. initiated this work. Y.F. and H.K. constructed the theory and carried out the calculations. Y.F., K.B., and A.K. wrote the paper.


\bibliography{QTuring}

\begin{thebibliography}{36}%
\makeatletter
\providecommand \@ifxundefined [1]{%
 \@ifx{#1\undefined}
}%
\providecommand \@ifnum [1]{%
 \ifnum #1\expandafter \@firstoftwo
 \else \expandafter \@secondoftwo
 \fi
}%
\providecommand \@ifx [1]{%
 \ifx #1\expandafter \@firstoftwo
 \else \expandafter \@secondoftwo
 \fi
}%
\providecommand \natexlab [1]{#1}%
\providecommand \enquote  [1]{``#1''}%
\providecommand \bibnamefont  [1]{#1}%
\providecommand \bibfnamefont [1]{#1}%
\providecommand \citenamefont [1]{#1}%
\providecommand \href@noop [0]{\@secondoftwo}%
\providecommand \href [0]{\begingroup \@sanitize@url \@href}%
\providecommand \@href[1]{\@@startlink{#1}\@@href}%
\providecommand \@@href[1]{\endgroup#1\@@endlink}%
\providecommand \@sanitize@url [0]{\catcode `\\12\catcode `\$12\catcode
  `\&12\catcode `\#12\catcode `\^12\catcode `\_12\catcode `\%12\relax}%
\providecommand \@@startlink[1]{}%
\providecommand \@@endlink[0]{}%
\providecommand \url  [0]{\begingroup\@sanitize@url \@url }%
\providecommand \@url [1]{\endgroup\@href {#1}{\urlprefix }}%
\providecommand \urlprefix  [0]{URL }%
\providecommand \Eprint [0]{\href }%
\providecommand \doibase [0]{https://doi.org/}%
\providecommand \selectlanguage [0]{\@gobble}%
\providecommand \bibinfo  [0]{\@secondoftwo}%
\providecommand \bibfield  [0]{\@secondoftwo}%
\providecommand \translation [1]{[#1]}%
\providecommand \BibitemOpen [0]{}%
\providecommand \bibitemStop [0]{}%
\providecommand \bibitemNoStop [0]{.\EOS\space}%
\providecommand \EOS [0]{\spacefactor3000\relax}%
\providecommand \BibitemShut  [1]{\csname bibitem#1\endcsname}%
\let\auto@bib@innerbib\@empty
\bibitem [{\citenamefont {Turing}(1952)}]{Turing1952}%
  \BibitemOpen
  \bibfield  {author} {\bibinfo {author} {\bibfnamefont {A.~M.}\ \bibnamefont
  {Turing}},\ }\bibfield  {title} {\bibinfo {title} {The chemical basis of
  morphogenesis},\ }\href@noop {} {\bibfield  {journal} {\bibinfo  {journal}
  {Philosophical Transactions of the Royal Society of London. Series B,
  Biological Sciences}\ }\textbf {\bibinfo {volume} {237}},\ \bibinfo {pages}
  {37} (\bibinfo {year} {1952})}\BibitemShut {NoStop}%
\bibitem [{\citenamefont {Meinhardt}(1982)}]{Meinhardt_book}%
  \BibitemOpen
  \bibfield  {author} {\bibinfo {author} {\bibfnamefont {H.}~\bibnamefont
  {Meinhardt}},\ }\href {https://books.google.co.jp/books?id=sLMTAQAAIAAJ}
  {\emph {\bibinfo {title} {Models of Biological Pattern Formation}}}\
  (\bibinfo  {publisher} {Academic Press},\ \bibinfo {year} {1982})\BibitemShut
  {NoStop}%
\bibitem [{\citenamefont {Koch}\ and\ \citenamefont
  {Meinhardt}(1994)}]{Koch1994}%
  \BibitemOpen
  \bibfield  {author} {\bibinfo {author} {\bibfnamefont {A.~J.}\ \bibnamefont
  {Koch}}\ and\ \bibinfo {author} {\bibfnamefont {H.}~\bibnamefont
  {Meinhardt}},\ }\bibfield  {title} {\bibinfo {title} {Biological pattern
  formation: from basic mechanisms to complex structures},\ }\href
  {https://doi.org/10.1103/RevModPhys.66.1481} {\bibfield  {journal} {\bibinfo
  {journal} {Rev. Mod. Phys.}\ }\textbf {\bibinfo {volume} {66}},\ \bibinfo
  {pages} {1481} (\bibinfo {year} {1994})}\BibitemShut {NoStop}%
\bibitem [{\citenamefont {Murray}(2002)}]{Murray_book}%
  \BibitemOpen
  \bibfield  {author} {\bibinfo {author} {\bibfnamefont {J.~D.}\ \bibnamefont
  {Murray}},\ }\href@noop {} {\emph {\bibinfo {title} {Mathematical
  Biology}}},\ Interdisciplinary Applied Mathematics\ (\bibinfo  {publisher}
  {Springer New York},\ \bibinfo {year} {2002})\BibitemShut {NoStop}%
\bibitem [{\citenamefont {Meinhardt}(2009)}]{Meinhardt_book2}%
  \BibitemOpen
  \bibfield  {author} {\bibinfo {author} {\bibfnamefont {H.}~\bibnamefont
  {Meinhardt}},\ }\href@noop {} {\emph {\bibinfo {title} {The Algorithmic
  Beauty of Sea Shells}}}\ (\bibinfo  {publisher} {Springer-Verlag Berlin
  Heidelberg},\ \bibinfo {year} {2009})\BibitemShut {NoStop}%
\bibitem [{\citenamefont {Kondo}\ and\ \citenamefont {Asai}(1995)}]{Kondo1995}%
  \BibitemOpen
  \bibfield  {author} {\bibinfo {author} {\bibfnamefont {S.}~\bibnamefont
  {Kondo}}\ and\ \bibinfo {author} {\bibfnamefont {R.}~\bibnamefont {Asai}},\
  }\bibfield  {title} {\bibinfo {title} {A reaction--diffusion wave on the skin
  of the marine angelfish pomacanthus},\ }\href
  {https://doi.org/10.1038/376765a0} {\bibfield  {journal} {\bibinfo  {journal}
  {Nature}\ }\textbf {\bibinfo {volume} {376}},\ \bibinfo {pages} {765}
  (\bibinfo {year} {1995})}\BibitemShut {NoStop}%
\bibitem [{\citenamefont {Kondo}\ and\ \citenamefont
  {Miura}(2010)}]{Kondo2010}%
  \BibitemOpen
  \bibfield  {author} {\bibinfo {author} {\bibfnamefont {S.}~\bibnamefont
  {Kondo}}\ and\ \bibinfo {author} {\bibfnamefont {T.}~\bibnamefont {Miura}},\
  }\bibfield  {title} {\bibinfo {title} {Reaction-diffusion model as a
  framework for understanding biological pattern formation},\ }\href@noop {}
  {\bibfield  {journal} {\bibinfo  {journal} {Science}\ }\textbf {\bibinfo
  {volume} {329}},\ \bibinfo {pages} {1616} (\bibinfo {year}
  {2010})}\BibitemShut {NoStop}%
\bibitem [{\citenamefont {Castets}\ \emph {et~al.}(1990)\citenamefont
  {Castets}, \citenamefont {Dulos}, \citenamefont {Boissonade},\ and\
  \citenamefont {De~Kepper}}]{Castets1990}%
  \BibitemOpen
  \bibfield  {author} {\bibinfo {author} {\bibfnamefont {V.}~\bibnamefont
  {Castets}}, \bibinfo {author} {\bibfnamefont {E.}~\bibnamefont {Dulos}},
  \bibinfo {author} {\bibfnamefont {J.}~\bibnamefont {Boissonade}},\ and\
  \bibinfo {author} {\bibfnamefont {P.}~\bibnamefont {De~Kepper}},\ }\bibfield
  {title} {\bibinfo {title} {Experimental evidence of a sustained standing
  turing-type nonequilibrium chemical pattern},\ }\href
  {https://doi.org/10.1103/PhysRevLett.64.2953} {\bibfield  {journal} {\bibinfo
   {journal} {Phys. Rev. Lett.}\ }\textbf {\bibinfo {volume} {64}},\ \bibinfo
  {pages} {2953} (\bibinfo {year} {1990})}\BibitemShut {NoStop}%
\bibitem [{\citenamefont {Ouyang}\ and\ \citenamefont
  {Swinney}(1991)}]{Ouyang1991}%
  \BibitemOpen
  \bibfield  {author} {\bibinfo {author} {\bibfnamefont {Q.}~\bibnamefont
  {Ouyang}}\ and\ \bibinfo {author} {\bibfnamefont {H.~L.}\ \bibnamefont
  {Swinney}},\ }\bibfield  {title} {\bibinfo {title} {Transition from a uniform
  state to hexagonal and striped turing patterns},\ }\href
  {https://doi.org/10.1038/352610a0} {\bibfield  {journal} {\bibinfo  {journal}
  {Nature}\ }\textbf {\bibinfo {volume} {352}},\ \bibinfo {pages} {610}
  (\bibinfo {year} {1991})}\BibitemShut {NoStop}%
\bibitem [{\citenamefont {Horv{\'a}th}\ \emph {et~al.}(2009)\citenamefont
  {Horv{\'a}th}, \citenamefont {Szalai},\ and\ \citenamefont
  {De~Kepper}}]{Horvath2009}%
  \BibitemOpen
  \bibfield  {author} {\bibinfo {author} {\bibfnamefont {J.}~\bibnamefont
  {Horv{\'a}th}}, \bibinfo {author} {\bibfnamefont {I.}~\bibnamefont
  {Szalai}},\ and\ \bibinfo {author} {\bibfnamefont {P.}~\bibnamefont
  {De~Kepper}},\ }\bibfield  {title} {\bibinfo {title} {An experimental design
  method leading to chemical turing patterns},\ }\href
  {https://doi.org/10.1126/science.1169973} {\bibfield  {journal} {\bibinfo
  {journal} {Science}\ }\textbf {\bibinfo {volume} {324}},\ \bibinfo {pages}
  {772} (\bibinfo {year} {2009})}\BibitemShut {NoStop}%
\bibitem [{\citenamefont {Bauer}(1958)}]{Bauer1958}%
  \BibitemOpen
  \bibfield  {author} {\bibinfo {author} {\bibfnamefont {E.}~\bibnamefont
  {Bauer}},\ }\bibfield  {title} {\bibinfo {title} {Ph{\"a}nomenologische
  theorie der kristallabscheidung an oberfl{\"a}chen. i},\ }\href
  {https://doi.org/https://doi.org/10.1524/zkri.1958.110.16.372} {\bibfield
  {journal} {\bibinfo  {journal} {Zeitschrift f{\"u}r Kristallographie -
  Crystalline Materials}\ }\textbf {\bibinfo {volume} {110}},\ \bibinfo {pages}
  {372 } (\bibinfo {year} {1958})}\BibitemShut {NoStop}%
\bibitem [{\citenamefont {Volmer}\ and\ \citenamefont
  {Weber}(1926)}]{Volmer1926}%
  \BibitemOpen
  \bibfield  {author} {\bibinfo {author} {\bibfnamefont {M.}~\bibnamefont
  {Volmer}}\ and\ \bibinfo {author} {\bibfnamefont {A.}~\bibnamefont {Weber}},\
  }\bibfield  {title} {\bibinfo {title} {Keimbildung in \"ubers\"attigten
  gebilden},\ }\href {https://doi.org/https://doi.org/10.1515/zpch-1926-11927}
  {\bibfield  {journal} {\bibinfo  {journal} {Zeitschrift f\"ur Physikalische
  Chemie}\ }\textbf {\bibinfo {volume} {119U}},\ \bibinfo {pages} {277 }
  (\bibinfo {year} {01 Jan. 1926})}\BibitemShut {NoStop}%
\bibitem [{\citenamefont {Frank}\ \emph {et~al.}(1949)\citenamefont {Frank},
  \citenamefont {van~der Merwe},\ and\ \citenamefont {Mott}}]{Frank1949}%
  \BibitemOpen
  \bibfield  {author} {\bibinfo {author} {\bibfnamefont {F.~C.}\ \bibnamefont
  {Frank}}, \bibinfo {author} {\bibfnamefont {J.~H.}\ \bibnamefont {van~der
  Merwe}},\ and\ \bibinfo {author} {\bibfnamefont {N.~F.}\ \bibnamefont
  {Mott}},\ }\bibfield  {title} {\bibinfo {title} {One-dimensional
  dislocations. i. static theory},\ }\href
  {https://doi.org/10.1098/rspa.1949.0095} {\bibfield  {journal} {\bibinfo
  {journal} {Proceedings of the Royal Society of London. Series A. Mathematical
  and Physical Sciences}\ }\textbf {\bibinfo {volume} {198}},\ \bibinfo {pages}
  {205} (\bibinfo {year} {1949})}\BibitemShut {NoStop}%
\bibitem [{\citenamefont {Stranski}\ and\ \citenamefont
  {Krastanow}(1937)}]{Stranski1937}%
  \BibitemOpen
  \bibfield  {author} {\bibinfo {author} {\bibfnamefont {I.~N.}\ \bibnamefont
  {Stranski}}\ and\ \bibinfo {author} {\bibfnamefont {L.}~\bibnamefont
  {Krastanow}},\ }\bibfield  {title} {\bibinfo {title} {Zur theorie der
  orientierten ausscheidung von ionenkristallen aufeinander},\ }\href
  {https://doi.org/10.1007/BF01798103} {\bibfield  {journal} {\bibinfo
  {journal} {Monatshefte f{\"u}r Chemie und verwandte Teile anderer
  Wissenschaften}\ }\textbf {\bibinfo {volume} {71}},\ \bibinfo {pages} {351}
  (\bibinfo {year} {1937})}\BibitemShut {NoStop}%
\bibitem [{\citenamefont {Fang}\ \emph {et~al.}(2018)\citenamefont {Fang},
  \citenamefont {Adamo}, \citenamefont {Jia}, \citenamefont {Cava},
  \citenamefont {Wu}, \citenamefont {Felser},\ and\ \citenamefont
  {Kapitulnik}}]{Fang2018}%
  \BibitemOpen
  \bibfield  {author} {\bibinfo {author} {\bibfnamefont {A.}~\bibnamefont
  {Fang}}, \bibinfo {author} {\bibfnamefont {C.}~\bibnamefont {Adamo}},
  \bibinfo {author} {\bibfnamefont {S.}~\bibnamefont {Jia}}, \bibinfo {author}
  {\bibfnamefont {R.~J.}\ \bibnamefont {Cava}}, \bibinfo {author}
  {\bibfnamefont {S.-C.}\ \bibnamefont {Wu}}, \bibinfo {author} {\bibfnamefont
  {C.}~\bibnamefont {Felser}},\ and\ \bibinfo {author} {\bibfnamefont
  {A.}~\bibnamefont {Kapitulnik}},\ }\bibfield  {title} {\bibinfo {title}
  {Bursting at the seams: Rippled monolayer bismuth on nbse2},\ }\bibfield
  {journal} {\bibinfo  {journal} {Science Advances}\ }\textbf {\bibinfo
  {volume} {4}},\ \href {https://doi.org/10.1126/sciadv.aaq0330}
  {10.1126/sciadv.aaq0330} (\bibinfo {year} {2018})\BibitemShut {NoStop}%
\bibitem [{\citenamefont {Hasan}\ and\ \citenamefont {Kane}(2010)}]{Hasan2010}%
  \BibitemOpen
  \bibfield  {author} {\bibinfo {author} {\bibfnamefont {M.~Z.}\ \bibnamefont
  {Hasan}}\ and\ \bibinfo {author} {\bibfnamefont {C.~L.}\ \bibnamefont
  {Kane}},\ }\bibfield  {title} {\bibinfo {title} {Colloquium: Topological
  insulators},\ }\href {https://doi.org/10.1103/RevModPhys.82.3045} {\bibfield
  {journal} {\bibinfo  {journal} {Rev. Mod. Phys.}\ }\textbf {\bibinfo {volume}
  {82}},\ \bibinfo {pages} {3045} (\bibinfo {year} {2010})}\BibitemShut
  {NoStop}%
\bibitem [{\citenamefont {Qi}\ and\ \citenamefont {Zhang}(2011)}]{Qi2011}%
  \BibitemOpen
  \bibfield  {author} {\bibinfo {author} {\bibfnamefont {X.-L.}\ \bibnamefont
  {Qi}}\ and\ \bibinfo {author} {\bibfnamefont {S.-C.}\ \bibnamefont {Zhang}},\
  }\bibfield  {title} {\bibinfo {title} {Topological insulators and
  superconductors},\ }\href {https://doi.org/10.1103/RevModPhys.83.1057}
  {\bibfield  {journal} {\bibinfo  {journal} {Rev. Mod. Phys.}\ }\textbf
  {\bibinfo {volume} {83}},\ \bibinfo {pages} {1057} (\bibinfo {year}
  {2011})}\BibitemShut {NoStop}%
\bibitem [{\citenamefont {Reis}\ \emph {et~al.}(2017)\citenamefont {Reis},
  \citenamefont {Li}, \citenamefont {Dudy}, \citenamefont {Bauernfeind},
  \citenamefont {Glass}, \citenamefont {Hanke}, \citenamefont {Thomale},
  \citenamefont {Sch{\"a}fer},\ and\ \citenamefont {Claessen}}]{Reis2017}%
  \BibitemOpen
  \bibfield  {author} {\bibinfo {author} {\bibfnamefont {F.}~\bibnamefont
  {Reis}}, \bibinfo {author} {\bibfnamefont {G.}~\bibnamefont {Li}}, \bibinfo
  {author} {\bibfnamefont {L.}~\bibnamefont {Dudy}}, \bibinfo {author}
  {\bibfnamefont {M.}~\bibnamefont {Bauernfeind}}, \bibinfo {author}
  {\bibfnamefont {S.}~\bibnamefont {Glass}}, \bibinfo {author} {\bibfnamefont
  {W.}~\bibnamefont {Hanke}}, \bibinfo {author} {\bibfnamefont
  {R.}~\bibnamefont {Thomale}}, \bibinfo {author} {\bibfnamefont
  {J.}~\bibnamefont {Sch{\"a}fer}},\ and\ \bibinfo {author} {\bibfnamefont
  {R.}~\bibnamefont {Claessen}},\ }\bibfield  {title} {\bibinfo {title}
  {Bismuthene on a sic substrate: A candidate for a high-temperature quantum
  spin hall material},\ }\href {https://doi.org/10.1126/science.aai8142}
  {\bibfield  {journal} {\bibinfo  {journal} {Science}\ }\textbf {\bibinfo
  {volume} {357}},\ \bibinfo {pages} {287} (\bibinfo {year}
  {2017})}\BibitemShut {NoStop}%
\bibitem [{\citenamefont {Chandrasekhar}(1981)}]{Chandrasekhar_book}%
  \BibitemOpen
  \bibfield  {author} {\bibinfo {author} {\bibfnamefont {S.}~\bibnamefont
  {Chandrasekhar}},\ }\href {https://books.google.co.jp/books?id=oU\_-6ikmidoC}
  {\emph {\bibinfo {title} {Hydrodynamic and Hydromagnetic Stability}}},\ Dover
  Books on Physics Series\ (\bibinfo  {publisher} {Dover Publications},\
  \bibinfo {year} {1981})\BibitemShut {NoStop}%
\bibitem [{\citenamefont {Dyson}\ and\ \citenamefont
  {Lenard}(1967)}]{Dyson1967}%
  \BibitemOpen
  \bibfield  {author} {\bibinfo {author} {\bibfnamefont {F.~J.}\ \bibnamefont
  {Dyson}}\ and\ \bibinfo {author} {\bibfnamefont {A.}~\bibnamefont {Lenard}},\
  }\bibfield  {title} {\bibinfo {title} {Stability of matter. i},\ }\href
  {https://doi.org/10.1063/1.1705209} {\bibfield  {journal} {\bibinfo
  {journal} {Journal of Mathematical Physics}\ }\textbf {\bibinfo {volume}
  {8}},\ \bibinfo {pages} {423} (\bibinfo {year} {1967})}\BibitemShut {NoStop}%
\bibitem [{\citenamefont {Littlewood}(1980)}]{Littlewood1980}%
  \BibitemOpen
  \bibfield  {author} {\bibinfo {author} {\bibfnamefont {P.~B.}\ \bibnamefont
  {Littlewood}},\ }\bibfield  {title} {\bibinfo {title} {The crystal structure
  of {IV}-{VI} compounds. i. classification and description},\ }\href
  {https://doi.org/10.1088/0022-3719/13/26/009} {\bibfield  {journal} {\bibinfo
   {journal} {Journal of Physics C: Solid State Physics}\ }\textbf {\bibinfo
  {volume} {13}},\ \bibinfo {pages} {4855} (\bibinfo {year}
  {1980})}\BibitemShut {NoStop}%
\bibitem [{\citenamefont {Behnia}(2016)}]{Behnia2016}%
  \BibitemOpen
  \bibfield  {author} {\bibinfo {author} {\bibfnamefont {K.}~\bibnamefont
  {Behnia}},\ }\bibfield  {title} {\bibinfo {title} {Finding merit in dividing
  neighbors},\ }\href {https://doi.org/10.1126/science.aad8688} {\bibfield
  {journal} {\bibinfo  {journal} {Science}\ }\textbf {\bibinfo {volume}
  {351}},\ \bibinfo {pages} {124} (\bibinfo {year} {2016})}\BibitemShut
  {NoStop}%
\bibitem [{\citenamefont {Tersoff}(1988)}]{Tersoff1988}%
  \BibitemOpen
  \bibfield  {author} {\bibinfo {author} {\bibfnamefont {J.}~\bibnamefont
  {Tersoff}},\ }\bibfield  {title} {\bibinfo {title} {New empirical approach
  for the structure and energy of covalent systems},\ }\href
  {https://doi.org/10.1103/PhysRevB.37.6991} {\bibfield  {journal} {\bibinfo
  {journal} {Phys. Rev. B}\ }\textbf {\bibinfo {volume} {37}},\ \bibinfo
  {pages} {6991} (\bibinfo {year} {1988})}\BibitemShut {NoStop}%
\bibitem [{SM()}]{SM}%
  \BibitemOpen
  \href@noop {} {\bibinfo {title} {Materials and methods are available as
  supplementary materials.}}\BibitemShut {Stop}%
\bibitem [{\citenamefont {Peierls}(1991)}]{Peierls_book2}%
  \BibitemOpen
  \bibfield  {author} {\bibinfo {author} {\bibfnamefont {R.}~\bibnamefont
  {Peierls}},\ }\href@noop {} {\emph {\bibinfo {title} {More Surprises in
  Theoretical Physics}}}\ (\bibinfo  {publisher} {Princeton University Press},\
  \bibinfo {year} {1991})\BibitemShut {NoStop}%
\bibitem [{\citenamefont {Hoffmann}(1988)}]{Hoffmann_book}%
  \BibitemOpen
  \bibfield  {author} {\bibinfo {author} {\bibfnamefont {R.}~\bibnamefont
  {Hoffmann}},\ }\href@noop {} {\emph {\bibinfo {title} {Solids and Surfaces: A
  Chemist's View of Bonding in Extended Structures}}}\ (\bibinfo  {publisher}
  {WILEY-VCH Verlag},\ \bibinfo {year} {1988})\BibitemShut {NoStop}%
\bibitem [{\citenamefont {Fuseya}\ \emph {et~al.}(2015)\citenamefont {Fuseya},
  \citenamefont {Ogata},\ and\ \citenamefont {Fukuyama}}]{Fuseya2015}%
  \BibitemOpen
  \bibfield  {author} {\bibinfo {author} {\bibfnamefont {Y.}~\bibnamefont
  {Fuseya}}, \bibinfo {author} {\bibfnamefont {M.}~\bibnamefont {Ogata}},\ and\
  \bibinfo {author} {\bibfnamefont {H.}~\bibnamefont {Fukuyama}},\ }\bibfield
  {title} {\bibinfo {title} {Transport properties and diamagnetism of dirac
  electrons in bismuth},\ }\href@noop {} {\bibfield  {journal} {\bibinfo
  {journal} {J. Phys. Soc. Jpn.}\ }\textbf {\bibinfo {volume} {84}},\ \bibinfo
  {pages} {012001} (\bibinfo {year} {2015})}\BibitemShut {NoStop}%
\bibitem [{\citenamefont {Young}(1975)}]{Young1975}%
  \BibitemOpen
  \bibfield  {author} {\bibinfo {author} {\bibfnamefont {D.~A.}\ \bibnamefont
  {Young}},\ }\href
  {http://inis.iaea.org/search/search.aspx?orig_q=RN:07255152} {\emph {\bibinfo
  {title} {Phase diagrams of the elements}}},\ \bibinfo {type} {Tech. Rep.}\
  (\bibinfo  {institution} {California Univ., Livermore (USA). Lawrence
  Livermore Lab},\ \bibinfo {address} {United States},\ \bibinfo {year}
  {1975})\BibitemShut {NoStop}%
\bibitem [{\citenamefont {Nagao}\ \emph {et~al.}(2004)\citenamefont {Nagao},
  \citenamefont {Sadowski}, \citenamefont {Saito}, \citenamefont {Yaginuma},
  \citenamefont {Fujikawa}, \citenamefont {Kogure}, \citenamefont {Ohno},
  \citenamefont {Hasegawa}, \citenamefont {Hasegawa},\ and\ \citenamefont
  {Sakurai}}]{Nagao2004}%
  \BibitemOpen
  \bibfield  {author} {\bibinfo {author} {\bibfnamefont {T.}~\bibnamefont
  {Nagao}}, \bibinfo {author} {\bibfnamefont {J.~T.}\ \bibnamefont {Sadowski}},
  \bibinfo {author} {\bibfnamefont {M.}~\bibnamefont {Saito}}, \bibinfo
  {author} {\bibfnamefont {S.}~\bibnamefont {Yaginuma}}, \bibinfo {author}
  {\bibfnamefont {Y.}~\bibnamefont {Fujikawa}}, \bibinfo {author}
  {\bibfnamefont {T.}~\bibnamefont {Kogure}}, \bibinfo {author} {\bibfnamefont
  {T.}~\bibnamefont {Ohno}}, \bibinfo {author} {\bibfnamefont {Y.}~\bibnamefont
  {Hasegawa}}, \bibinfo {author} {\bibfnamefont {S.}~\bibnamefont {Hasegawa}},\
  and\ \bibinfo {author} {\bibfnamefont {T.}~\bibnamefont {Sakurai}},\
  }\bibfield  {title} {\bibinfo {title} {Nanofilm allotrope and phase
  transformation of ultrathin bi film on si(111)-7x7},\ }\href
  {https://doi.org/10.1103/PhysRevLett.93.105501} {\bibfield  {journal}
  {\bibinfo  {journal} {Phys. Rev. Lett.}\ }\textbf {\bibinfo {volume} {93}},\
  \bibinfo {pages} {105501} (\bibinfo {year} {2004})}\BibitemShut {NoStop}%
\bibitem [{\citenamefont {Bollmann}\ \emph {et~al.}(2011)\citenamefont
  {Bollmann}, \citenamefont {van Gastel}, \citenamefont {Zandvliet},\ and\
  \citenamefont {Poelsema}}]{Bollmann2011}%
  \BibitemOpen
  \bibfield  {author} {\bibinfo {author} {\bibfnamefont {T.~R.~J.}\
  \bibnamefont {Bollmann}}, \bibinfo {author} {\bibfnamefont {R.}~\bibnamefont
  {van Gastel}}, \bibinfo {author} {\bibfnamefont {H.~J.~W.}\ \bibnamefont
  {Zandvliet}},\ and\ \bibinfo {author} {\bibfnamefont {B.}~\bibnamefont
  {Poelsema}},\ }\bibfield  {title} {\bibinfo {title} {Quantum size effect
  driven structure modifications of bi films on ni(111)},\ }\href
  {https://doi.org/10.1103/PhysRevLett.107.176102} {\bibfield  {journal}
  {\bibinfo  {journal} {Phys. Rev. Lett.}\ }\textbf {\bibinfo {volume} {107}},\
  \bibinfo {pages} {176102} (\bibinfo {year} {2011})}\BibitemShut {NoStop}%
\bibitem [{\citenamefont {Kudo}\ \emph {et~al.}(2007)\citenamefont {Kudo},
  \citenamefont {Mino},\ and\ \citenamefont {Nakamura}}]{Kudo2007}%
  \BibitemOpen
  \bibfield  {author} {\bibinfo {author} {\bibfnamefont {K.}~\bibnamefont
  {Kudo}}, \bibinfo {author} {\bibfnamefont {M.}~\bibnamefont {Mino}},\ and\
  \bibinfo {author} {\bibfnamefont {K.}~\bibnamefont {Nakamura}},\ }\bibfield
  {title} {\bibinfo {title} {Magnetic domain patterns depending on the sweeping
  rate of magnetic fields},\ }\href {https://doi.org/10.1143/JPSJ.76.013002}
  {\bibfield  {journal} {\bibinfo  {journal} {J. Phys. Soc. Jpn.}\ }\textbf
  {\bibinfo {volume} {76}},\ \bibinfo {pages} {013002} (\bibinfo {year}
  {2007})}\BibitemShut {NoStop}%
\bibitem [{\citenamefont {Jeudy}\ and\ \citenamefont
  {Gourdon}(2006)}]{Jeudy2006}%
  \BibitemOpen
  \bibfield  {author} {\bibinfo {author} {\bibfnamefont {V.}~\bibnamefont
  {Jeudy}}\ and\ \bibinfo {author} {\bibfnamefont {C.}~\bibnamefont
  {Gourdon}},\ }\bibfield  {title} {\bibinfo {title} {Instability-driven
  formation of domains in the intermediate state of type-i superconductors},\
  }\href {https://doi.org/10.1209/epl/i2006-10123-8} {\bibfield  {journal}
  {\bibinfo  {journal} {Europhysics Letters ({EPL})}\ }\textbf {\bibinfo
  {volume} {75}},\ \bibinfo {pages} {482} (\bibinfo {year} {2006})}\BibitemShut
  {NoStop}%
\bibitem [{\citenamefont {Prozorov}(2007)}]{Prozorov2007}%
  \BibitemOpen
  \bibfield  {author} {\bibinfo {author} {\bibfnamefont {R.}~\bibnamefont
  {Prozorov}},\ }\bibfield  {title} {\bibinfo {title} {Equilibrium topology of
  the intermediate state in type-i superconductors of different shapes},\
  }\href {https://doi.org/10.1103/PhysRevLett.98.257001} {\bibfield  {journal}
  {\bibinfo  {journal} {Phys. Rev. Lett.}\ }\textbf {\bibinfo {volume} {98}},\
  \bibinfo {pages} {257001} (\bibinfo {year} {2007})}\BibitemShut {NoStop}%
\bibitem [{\citenamefont {Paulsen}\ \emph {et~al.}(2012)\citenamefont
  {Paulsen}, \citenamefont {Hykel}, \citenamefont {Hasselbach},\ and\
  \citenamefont {Aoki}}]{Paulsen2012}%
  \BibitemOpen
  \bibfield  {author} {\bibinfo {author} {\bibfnamefont {C.}~\bibnamefont
  {Paulsen}}, \bibinfo {author} {\bibfnamefont {D.~J.}\ \bibnamefont {Hykel}},
  \bibinfo {author} {\bibfnamefont {K.}~\bibnamefont {Hasselbach}},\ and\
  \bibinfo {author} {\bibfnamefont {D.}~\bibnamefont {Aoki}},\ }\bibfield
  {title} {\bibinfo {title} {Observation of the meissner-ochsenfeld effect and
  the absence of the meissner state in ucoge},\ }\href
  {https://doi.org/10.1103/PhysRevLett.109.237001} {\bibfield  {journal}
  {\bibinfo  {journal} {Phys. Rev. Lett.}\ }\textbf {\bibinfo {volume} {109}},\
  \bibinfo {pages} {237001} (\bibinfo {year} {2012})}\BibitemShut {NoStop}%
\bibitem [{\citenamefont {Bistritzer}\ and\ \citenamefont
  {MacDonald}(2011)}]{Bistritzer2011}%
  \BibitemOpen
  \bibfield  {author} {\bibinfo {author} {\bibfnamefont {R.}~\bibnamefont
  {Bistritzer}}\ and\ \bibinfo {author} {\bibfnamefont {A.~H.}\ \bibnamefont
  {MacDonald}},\ }\bibfield  {title} {\bibinfo {title} {Moir{\'e} bands in
  twisted double-layer graphene},\ }\href
  {https://doi.org/10.1073/pnas.1108174108} {\bibfield  {journal} {\bibinfo
  {journal} {Proceedings of the National Academy of Sciences}\ }\textbf
  {\bibinfo {volume} {108}},\ \bibinfo {pages} {12233} (\bibinfo {year}
  {2011})}\BibitemShut {NoStop}%
\bibitem [{\citenamefont {Cao}\ \emph {et~al.}(2018)\citenamefont {Cao},
  \citenamefont {Fatemi}, \citenamefont {Fang}, \citenamefont {Watanabe},
  \citenamefont {Taniguchi}, \citenamefont {Kaxiras},\ and\ \citenamefont
  {Jarillo-Herrero}}]{Cao2018}%
  \BibitemOpen
  \bibfield  {author} {\bibinfo {author} {\bibfnamefont {Y.}~\bibnamefont
  {Cao}}, \bibinfo {author} {\bibfnamefont {V.}~\bibnamefont {Fatemi}},
  \bibinfo {author} {\bibfnamefont {S.}~\bibnamefont {Fang}}, \bibinfo {author}
  {\bibfnamefont {K.}~\bibnamefont {Watanabe}}, \bibinfo {author}
  {\bibfnamefont {T.}~\bibnamefont {Taniguchi}}, \bibinfo {author}
  {\bibfnamefont {E.}~\bibnamefont {Kaxiras}},\ and\ \bibinfo {author}
  {\bibfnamefont {P.}~\bibnamefont {Jarillo-Herrero}},\ }\bibfield  {title}
  {\bibinfo {title} {Unconventional superconductivity in magic-angle graphene
  superlattices},\ }\href {https://doi.org/10.1038/nature26160} {\bibfield
  {journal} {\bibinfo  {journal} {Nature}\ }\textbf {\bibinfo {volume} {556}},\
  \bibinfo {pages} {43} (\bibinfo {year} {2018})}\BibitemShut {NoStop}%
\end{thebibliography}%

\end{document}